\documentclass[aps,prd,twocolumn,superscriptaddress,nofootinbib]{revtex4-1}

\usepackage{graphicx}
\usepackage{booktabs}
\usepackage[hidelinks]{hyperref}
\usepackage{color}
\hypersetup{
    colorlinks,
    linkcolor={blue},
    citecolor={blue},
    urlcolor={blue}
}
\usepackage{amssymb,amsmath,amsfonts}
\usepackage[utf8]{inputenc}
\usepackage{nicefrac}
\usepackage{braket}
\usepackage{slashed}

\newcommand{\e}{\mathrm{e}}
\renewcommand{\d}{\mathrm d}
\renewcommand{\eqref}[1]{Eq.~(\ref{#1})}
\newcommand{\figref}[1]{Fig.~\ref{#1}}
\def\Nint{N_{\mathrm{int}}}
\def\Qmax{Q_{\mathrm{max}}}
\def\Qthr{Q_{\mathrm{thresh}}}
\def\Tc{T_{\mathrm{c}}}

\let\originalleft\left
\let\originalright\right
\renewcommand{\left}{\mathopen{}\mathclose\bgroup\originalleft}
\renewcommand{\right}{\aftergroup\egroup\originalright}

\begin{document}

\title{\boldmath
 $\chi_{\textrm{top}}(T \gg \Tc)$ in pure-glue QCD through reweighting
}

\author{P. Thomas Jahn}
\author{Guy D. Moore}
\author{Daniel Robaina}
\affiliation{Institut f\"ur Kernphysik (Theoriezentrum), Technische Universit\"at Darmstadt,\\ Schlossgartenstra{\ss}e 2, D-64289 Darmstadt, Germany}
\email{tjahn@theorie.ikp.physik.tu-darmstadt.de}
\email{guymoore@theorie.ikp.physik.tu-darmstadt.de}
\email{robaina@theorie.ikp.physik.tu-darmstadt.de}

\begin{abstract}
We calculate the topological susceptibility at $2.5~\Tc$ and
$4.1~\Tc$ in SU(3) pure Yang-Mills theory. We define topology
with the help of gradient flow and we largely overcome the problem of
poor statistics at high temperatures by applying a reweighting
technique in terms of the topological charge, measured after a
specific small amount of gradient flow. This allows us to
obtain a sample of configurations which compares topological sectors
with good statistics, with enhanced
tunneling between topologies. We quote continuum extrapolated results
at these two temperatures and conclude that our method is viable and
can be extended without new conceptual problems to the case of
full QCD with fermions.
\end{abstract}

\maketitle
\flushbottom

\section{Introduction \label{sec:Introduction}}

Two of the most challenging problems in particle physics are the
strong \emph{CP} problem and the origin of dark matter. The axion
\cite{Weinberg:1977ma,Wilczek:1977pj}, a hypothetical light scalar
particle which appears in the Peccei-Quinn mechanism
\cite{Peccei:1977hh,Peccei:1977ur}, could solve both problems at once:
The additional degrees of freedom explain that the \emph{CP} violating phase
$\Theta_\mathrm{QCD}$ in the QCD Lagrangian vanishes
\cite{Peccei:1977ur} and the corresponding particle could play the
role of dark matter in the Universe
\cite{Preskill:1982cy,Abbott:1982af,Dine:1982ah}.

There is currently a lot of experimental effort to detect axions; for
a review on this we refer to Ref.~\cite{Irastorza:2018dyq}. From a
theory point of view, the properties of the axion are sensitive to the
topological susceptibility of QCD $\chi(T)$, defined as
\begin{equation}
  \chi(T) = \int \d^4 x \left\langle q(x) q(0) \right\rangle_\beta
  = \frac{1}{\beta V} \left\langle Q^2 \right\rangle,
\end{equation}
with $q(x)$ the topological charge density and
$Q=\int \d^4 x \; q(x)$ its integral (defined below).  While the
topological susceptibility at
low temperatures is well established \cite{diCortona:2015ldu},
calculations become much more challenging at high temperatures, and
axion cosmology requires knowledge of $\chi(T)$ at temperatures up to
about $7~\Tc$ \cite{Moore:2017ond,Klaer:2017ond}.
Recently there has been a burst of progress in determining $\chi(T)$
at high temperatures
\cite{Frison:2016vuc,Berkowitz:2015aua,Taniguchi:2016tjc,%
Bonati:2015vqz,Petreczky:2016vrs,Borsanyi:2015cka,Borsanyi:2016ksw}.
However, we feel that it would still be valuable to make an independent
study of topological susceptibility which reaches up to 7 $\Tc$.

At high temperatures topology is expected to be dominated by rare
single instanton \cite{Belavin:1975fg,tHooft:1976rip}
(really caloron \cite{Harrington:1978ve}) configurations
with a typical size $\rho \sim 0.4/T$ \cite{Gross:1980br}.  These
configurations become more suppressed as one considers higher
temperatures, by $\chi(T) \propto T^{-7}$ (at lowest perturbative
order in pure-glue gauge theory; with light fermions there is an
additional factor of $T^{-N_\mathrm f/3}$).
This makes studying topology by lattice Monte Carlo simulations
challenging; in an ensemble of high-temperature gauge theory
configurations, virtually none of the configurations will possess
topology, leading to severely limited statistics.
For instance, if we keep the number of temporal points across the
lattice fixed, the instanton density in terms of lattice sites
vanishes as $T^{-11}$.  Furthermore, the efficiency with which a
Markov chain algorithm samples these topological configurations will
be additionally suppressed, because the chain must pass through
``small'' instantons (or dislocations) to move between distinct
topological sectors, and these dislocations get rarer with decreasing
lattice spacing as $a^{-11}$.

One way around this problem is to measure topology at a low
temperature where instantons are not rare, and to work up to high
temperatures differentially by studying fixed-topology ensembles
\cite{Borsanyi:2016ksw,Frison:2016vuc}.  But we feel it is important
as a cross-check to be able to perform a direct study of topology at
high temperature.  This will require a reweighting procedure to
overcome the sampling challenges.  Our goal in this paper is to
present such a reweighting approach.  Since this work is exploratory,
we will content ourselves with a study of the quenched (or pure-glue)
theory.  Once the technique is well established, we see no obstacles
in adapting it to the unquenched case (though there will be the usual
increase in numerical cost associated with going from pure glue to
unquenched).

This paper is organized as follows: In Sec.~\ref{sec: Method}, we
discuss in detail the method that we use to enhance the number of
instantons in the lattice simulations, namely a combination of
gradient flow and reweighting. Results of the topological
susceptibility of the quenched theory up to $4.1~\Tc$ obtained
with this method are presented in Sec.~\ref{sec: Results}. Conclusions
and a discussion can then be found in Sec.~\ref{sec: Discussion}.

\section{Method \label{sec: Method}}

The statistical problem of calculating the topological susceptibility
at high temperatures on the lattice is twofold. First, the
quantity is expected to be physically very small which will result in
most of the configurations having $Q=0$. Only by collecting a huge
amount of statistics, a meaningful statement about expectation values
involving the topological charge can be made. This problem becomes
more severe as one goes to higher temperatures.  And in addition, the
algorithm itself used to generate the sample (usually
heatbath/overrelaxation or HMC) tends to get stuck in the different
topological sectors, with tunneling events between sectors more and
more suppressed as one takes the continuum limit.  Here we show
how to use reweighting to overcome both problems.

\subsection{Definition of reweighting}

Our reweighting approach is an evolution of those in Refs.
\cite{Berg:1991cf,Kajantie:1995kf,Laine:1998qk,Wang:2000fzi}. Since
the topological charge is not restricted to integer values on the lattice,
rare events exist that
will enable tunneling between different sectors. The goal is then to
enhance those and use them to generate a sample of configurations
almost homogeneously distributed across the topological sectors of
interest. At the same time, it is mandatory to be able to know by how
much they were enhanced so that this effect can be removed at the end
without losing the statistical power. In the following, we describe
one way of achieving this for the case of pure SU(3) Yang-Mills theory
with periodic boundary conditions.

The nonperturbative approach of lattice gauge theory is based on a
stochastic evaluation of the partition function
\begin{equation}
\mathcal{Z} = \int \mathcal{D}U \e^{-S_\mathrm{W}[U]}.
\label{eq:part}
\end{equation}
Here $S_\mathrm{W}[U]$ is the ordinary SU(3) plaquette Wilson action
and $U$ are the links. The algorithmic challenge consists in obtaining a
sample of configurations which is precisely distributed according to the
probability distribution 
\begin{equation}
\d P(U) = \frac{1}{\mathcal Z} \mathrm \e^{-S_\mathrm{W}[U]}\d U.
\label{eq:prob_distr1}
\end{equation}
In this way, importance sampling enables the calculation of expectation
values of gauge invariant operators via the simple mean
\begin{equation}
\langle O \rangle = \frac{1}{N}\sum^N_i \mathcal{O}_i.
\label{eq:norew_exp}
\end{equation}
At high temperatures well above $\Tc$, the ordinary approach
just described will yield an ensemble with very little topological information.
Reweighting works by rewriting \eqref{eq:part} as
\begin{equation}
\mathcal{Z} = \int \mathcal{D}U \e^{-S_\mathrm{W}[U] + W(\xi)} \e^{-W(\xi)},
\end{equation}  
and therefore, \eqref{eq:norew_exp} turns into 
\begin{equation}
\langle \mathcal{O}\rangle = \frac{\sum^N_i \mathcal{O}_i\e^{-W(\xi_i)}}{\sum^N_i \e^{-W(\xi_i)}}
\label{eq:exp_rew}
\end{equation}
if the ensemble is distributed according to the modified probability distribution
\begin{equation}
\d P_\mathrm{rew}(U) = \frac{\e^{-S_\mathrm{W}[U]+W(\xi)}\d U}{\int \mathcal DU \e^{-S_\mathrm W[U] + W(\xi)}}.
\label{eq:prob_distr2}
\end{equation}
Notice that it is guaranteed that Eqs.~(\ref{eq:norew_exp}) and
(\ref{eq:exp_rew}) yield identical results for any choice of the
\emph{reweighting function} $W$ as long as our algorithm converges to
\eqref{eq:prob_distr2} and $N\to \infty$. The argument $\xi$ is an
arbitrary set of \emph{reweighting variables} which need to be
measured on each produced configuration.

\subsection{Choice of reweighting variable}
\label{subsec:choice}

If chosen correctly,
reweighting variables can account for a clear distinction between
different phases and therefore favor or suppress certain sectors in
Monte Carlo space. A natural choice is then the topological charge
itself:
\begin{equation}
  Q = \sum_x q(x)
  = \frac{1}{64 \pi^2} \epsilon_{\mu\nu\rho\sigma} \sum_x
  \hat{F}_{\mu\nu}^a(x) \hat{F}_{\rho\sigma}^a(x).
\label{Qdef}
\end{equation}
Here $\hat{F}_{\mu\nu}(x)$ is a lattice discretized form of the field
strength.  The conventional choice is the ``clover'' value (the
average over the four square plaquettes touching the point $x$), but
we use an $a^2$-improved choice composed of a linear combination of
squares and $1\times 2$ rectangles \cite{Jahn:2018jvx}, specifically
\begin{align}
\begin{split}
  \hat{F}_{\mathrm{clov}} &= 
  \frac{1}{4} \begin{picture}(25,20)
    \thicklines
    \multiput(4,-4)(0,8){3}{\line(1,0){16}}
    \multiput(4,-4)(8,0){3}{\line(0,1){16}}
    \put(12,4){\circle*{4}}
  \end{picture}
  \\
  \hat{F}_{\mathrm{imp}} &= \frac{5}{12} \begin{picture}(25,20)
    \thicklines
    \multiput(4,-4)(0,8){3}{\line(1,0){16}}
    \multiput(4,-4)(8,0){3}{\line(0,1){16}}
    \put(12,4){\circle*{4}}
  \end{picture}
  - \frac{1}{24} \left( \begin{picture}(40,20)
    \thicklines
    \multiput(4,-4)(0,8){3}{\line(1,0){32}}
    \multiput(4,-4)(16,0){3}{\line(0,1){16}}
    \multiput(12,-4)(16,0){2}{\multiput(0,0)(0,2){8}{\line(0,1){1}}}
    \put(20,4){\circle*{4}}
  \end{picture}
  +
  \begin{picture}(25,22)
    \thicklines
    \multiput(4,-12)(0,16){3}{\line(1,0){16}}
    \multiput(4,-12)(8,0){3}{\line(0,1){32}}
    \multiput(4,-4)(0,16){2}{\multiput(0,0)(2,0){8}{\line(1,0){1}}}
    \put(12,4){\circle*{4}}
  \end{picture} \right) .
  \end{split}
\end{align}

However, using $Q$ directly on the original configuration actually
fails, because the topological density contains
high-dimension operator corrections which are not topological and
which receive large random additive contributions.  The solution is
well-known; we should apply some amount of gradient flow
\cite{Narayanan:2006rf,Luscher:2009eq} to remove the UV fluctuations
responsible for this problem.  We therefore define our single
reweighting variable $\xi$ as
\begin{equation}
\xi =  Q' = \frac{1}{64 \pi^2} \epsilon_{\mu\nu\rho\sigma}\sum_{x}
\left( \hat F^a_{\mu\nu}(x) \hat F^a_{\rho\sigma}(x) \right)_{t'}
\label{eq:rew_var}
\end{equation}
where $t'$ denotes a relatively small amount of Wilson flow.
Specifically, we choose $t'$ to be enough Wilson flow that topology-1
configurations are clearly distinguished from random fluctuations,
\textsl{but not enough} to remove ``dislocations,'' small
concentrations of topological charge which are the intermediate steps
between the $Q=0$ and $Q=1$ sectors.  Therefore, $Q'$ is able to
distinguish between fluctuations about the $Q=0$ sector, dislocations
which lie between topological sectors, and genuine $Q=1$
configurations.  The true topological charge of the
configuration is denoted by $Q$ and is measured after a larger amount
of flow. We shall come back to this distinction in Sec.~\ref{subsec:totune}.

\subsection{Updating with an arbitrary weight function}
\label{subsec:algorithm}

Next we describe the Markov chain algorithm whose equilibrium
probability distribution is
\begin{equation}
\d P_\mathrm{rew}(U) = \frac{\e^{-S_\mathrm{W}[U]+W(Q')}\d U}{\int \mathcal DU \e^{-S_\mathrm W[U] + W(Q')}}
\end{equation}
assuming that the function $W(Q')$ is already known. Although it is common
practice to use the heatbath/overrelaxation algorithm in the context
of pure gauge theories, the hybrid Monte Carlo algorithm (HMC)
supports a conceptually simple fermionic extension, so we will use it
instead.

One of the simplest ways of producing a sample according to a given
probability distribution is to use a Metropolis-inspired
algorithm. This algorithm fulfills detailed balance and therefore the
Markov chain has an equilibrium distribution to which the system
converges if enough updates are done. After having evolved the
Hamilton equations as part of the standard molecular dynamics
evolution, an accept or reject step accepts the configuration with
probability
\begin{equation}
P_\mathrm{HMC} = \text{min}\left\{1, \e^{-\Delta H}\right\},
\end{equation}
where $\Delta H = H_\mathrm f - H_\mathrm i$ is the energy difference
(the subscripts ``i" and ``f" refer to ``initial" and ``final," respectively). It is given
by $H(\pi, U) = \frac{1}{2}\sum_x\pi(x)^2 + S_\mathrm{W}[U]$. This step
alone is of course not sufficient for incorporating reweighting. Therefore,
the configuration cannot be fully accepted yet. An additional reweighting
accept or reject step in terms of $Q'$ accepts the configuration with probability
\begin{equation}
P_\mathrm{rew} = \text{min}\left\{1,\e^{\Delta W}\right\},
\end{equation} 
where $\Delta W = W_\mathrm f - W_\mathrm i$. In total, the transition
probability $P(C_\mathrm i \rightarrow C_\mathrm f)$ is given by
\begin{align}
P(C_\mathrm i \rightarrow C_\mathrm f) = &\int \d\pi_\mathrm i
\d\pi_\mathrm f P_\mathrm G(\pi_i) P_\mathrm M((\pi_\mathrm i,U_\mathrm i)
\rightarrow (\pi_\mathrm f,U_\mathrm f)) \nonumber
\\
&\times P_\mathrm{HMC}(\Delta H) P_\mathrm{rew}(\Delta W).
\end{align}
The probability $P_\mathrm G(\pi_\mathrm i) \sim \e^{-\frac{1}{2}\pi^2_\mathrm i}$
with which the conjugate momenta are chosen is drawn Gaussian as usual.
The probability $P_\mathrm M$ refers to the molecular dynamics evolution
which is a deterministic process. Therefore, $P_\mathrm M$ can be seen as
a $\delta$-function that evolves the fields from $(\pi_\mathrm i,U_\mathrm i)
\rightarrow (\pi_\mathrm f,U_\mathrm f)$ with a unit probability.
Our procedure can be summarized as follows:

\begin{enumerate}
\item Generate a candidate configuration by evolving the Hamilton equations.
\item Perform a Metropolis step in terms of $\Delta H(\pi, U)$.
\begin{enumerate}
\item[2.1] If accepted, store the candidate configuration.
\item[2.2] If rejected, return to the initial current configuration (go to step 1).
\end{enumerate}
\item If 2.1 is true, integrate the flow equation up to flow time $t'$ and perform
a Metropolis step in terms of $\Delta W(Q')$.
\begin{enumerate}
\item[3.1] If accepted, return to the unflowed candidate configuration and fully accept it. 
\item[3.2] If rejected, return to the initial current configuration (go to step 1).
\end{enumerate}
\end{enumerate}

We have described the algorithm assuming a known reweighting function
$W(Q')$. In the next section we shall address the question of how
to choose $W(Q')$ such that the final sample is, in the best case
scenario, homogeneously distributed across topological sectors.

\subsection{\boldmath Building the reweighting function $W(Q')$}
\label{sec:build_W}

As explained in the previous subsection, the \emph{a priori} knowledge of
the reweighting function $W$ is mandatory to implement
reweighting. In this subsection we describe how to find an optimal
choice in a completely automated way.  Our approach is similar to
Refs.~\cite{Laine:1998qk,Wang:2000fzi}.

We perform two HMC Markov chains; one to determine $W$ and one to
apply the (now fixed) $W$ function to perform our actual Monte Carlo
study.  Here we describe the preparatory Markov chain which determines
$W$. This preparatory run consists of reweighting updates as described in
Sec.~\ref{subsec:algorithm} with the only difference that the function $W$
is updated after each trajectory. In this way, we are able to force
the system to visit certain sectors in Monte Carlo space that are rare
and, at the same time, avoid those that already were visited quite
often.

First, we need to define a \emph{reweighting domain}
$\Omega_\mathrm{rew}$. This is the interval in $Q'$ where $W$ will
account for reweighting. A natural choice is $(-\Qmax',
\Qmax')$, where $\Qmax'$ is the highest integer value
corresponding to the highest topological sector that we want to
include in the reweighting sample. Since we are ultimately
interested in $\langle Q^2 \rangle$, we can make use of the symmetry
$Q' \mapsto -Q'$ and redefine $Q'_\mathrm{new} = |Q'|$. For convenience, we drop the subscript ``new'' in what follows. In this way our reweighting domain is
\begin{equation}
  \Omega_\mathrm{rew} = \left[ 0,\Qmax' \right].
\end{equation}
We divide this domain into $\Nint$ intervals,
$0<Q'_1<Q'_2<...<Q_{\Nint}'=\Qmax'$, and we name the interval between
$Q_i'$ and $Q_{i+1}'$, $\omega_i$.
We define $W(Q')$ by
giving it definite values at each $Q_i'$ and interpolating linearly
between these values; that is, $W(Q')$ is taken as piecewise linear.
The last interval, $\omega_{\Nint}$, is all points with $Q'>\Qmax'$;
we choose $W\left(Q'>\Qmax'\right) = W(\Qmax')$ in this (semi-infinite) interval.
In other words, values above the top edge of our
reweighting domain are not rejected; they are just not reweighted any
higher than the boundary value of the domain.  To summarize,
\begin{equation}
  W(Q') =  \left\{   \begin{array}{ll}
  (1-x)W_i + xW_{i+1}, \quad & Q' \in \omega_i \\
    W_{\Nint}, & Q'>\Qmax' \\ \end{array}
  \right.
\end{equation}
with
\begin{equation}
x = \frac{Q'-Q'_i}{Q'_{i+1}-Q'_i}.
\end{equation}
\begin{figure}[t]
  \centering
  \includegraphics[width=1\linewidth]{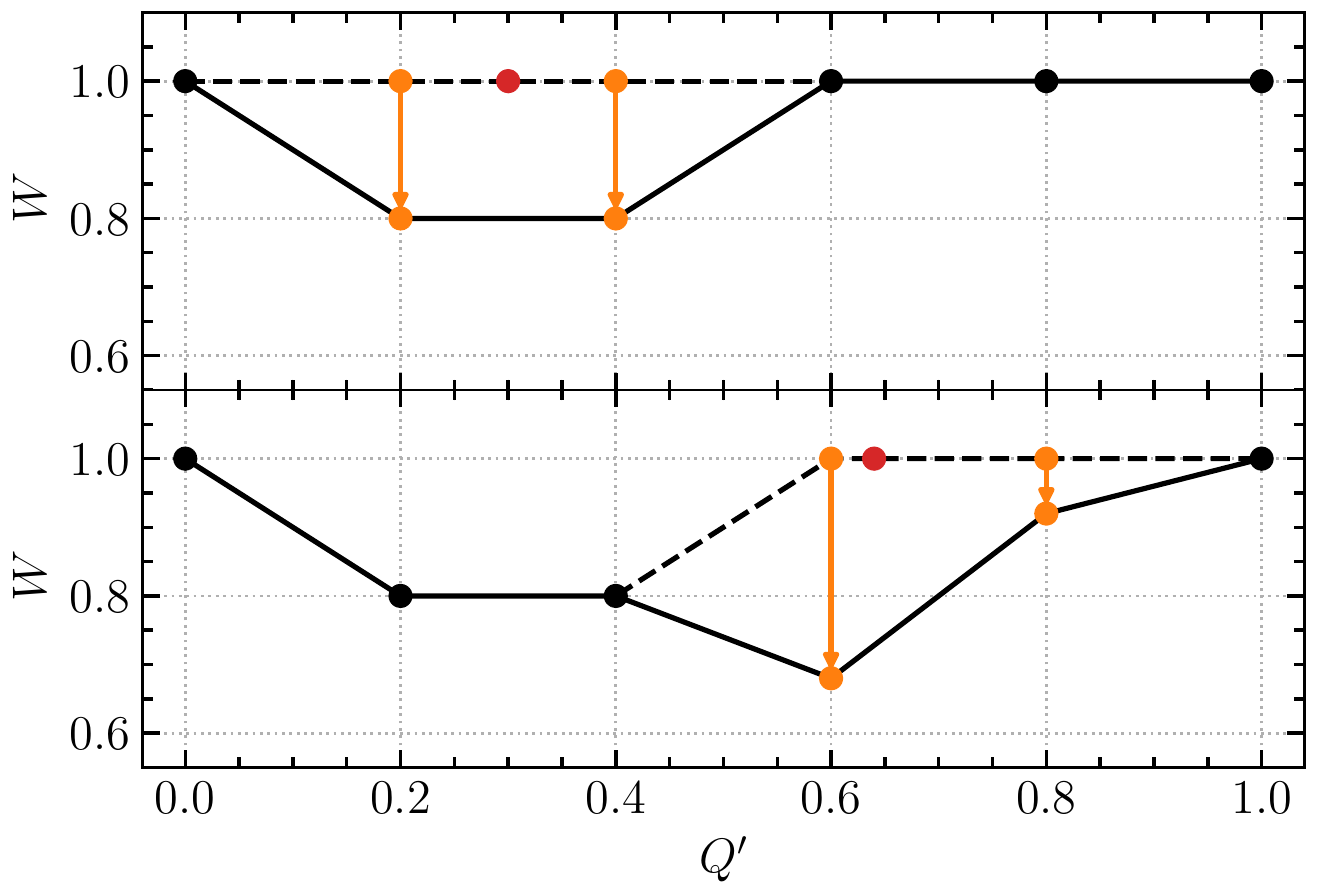}
  \caption{Schematic depiction of how $W$ is built with $N_\mathrm{int} = 5$ and $s=0.4$.
  The red dot indicates the measured $Q'$, while the orange points and arrows
  show the change in the $W$ function. The solid black line shows the updated $W$ function,
  the dashed line is the updated part of $W$ before the update.
  Top: $Q'=0.3$. Bottom: $Q'=0.64$.}
  \label{fig:W_scheme}
\end{figure}

Having defined our reweighting function in the domain of interest, we
can start making reweighting updates. After letting the system
thermalize with ordinary HMC updates, we begin building the
reweighting function with reweighting updates. We start with a
constant function $W(Q')\equiv 1$.\footnote{Notice that overall
  additive constants are irrelevant.}  Our philosophy is that,
whatever value of $Q'$ we currently have, this value is presumably
oversampled, and should be made less common by reducing $W(Q')$ at the
current value.  Because $W$ is piecewise linear, the most local change
we can make is to change the values at the two edges of the current
interval.  If $Q' \in \omega_i$ with $i\neq 0$, only
the corresponding values $W_i$ and $W_{i+1}$ are changed according to
\begin{align}
	W_i &\to W_i - s(1-x),
	\\
	W_{i+1} &\to W_{i+1} - sx,
\end{align}
while the rest of the function remains unaffected.  This procedure is
illustrated in \figref{fig:W_scheme}.

The first interval, $Q' \in \omega_0$, is a special case. We need to
remember that $Q'$ is strictly
positive and therefore $W_0$ will get updated less than the rest of
the points. We correct for this via
\begin{align}
	W_0 &\to W_0 -2s(1-x),
	\\
	W_1 &\to W_1 -sx.
\end{align}
The value of $s$ controls by how much $W$ changes each update. As soon
as the gross features of $W$ arise, we decrease its value to slowly
reach convergence. In order to do so it is instructive to introduce
the notion of \emph{complete sweep}. We refer to a complete sweep when
the reweighting variable $Q'$ ranges from $\omega_0$ to
$\omega_{\Nint-1}$ \emph{and} back to $\omega_0$.
We count the number of updates  needed to accomplish
this and name it $M$.  There is no need that it visits all intervals.
The combination $\delta W = sM/\Nint$ tells us how much in
average one point of $W$ has been changed during the completion
of the last sweep. After each complete sweep we compute $\delta W$
and reduce $s$ to $s\to\text{max}\left\{s/2, s\left(1 - \frac{\delta W}{1.5 \times 2}\right)\right\}$.
Therefore a sweep which changes a point in $W$ by of average more
than 1.5 will lead to $s$ being cut in half, while a sweep which changes
a point in $W$ by a smaller average amount will result in reducing $s$ less.
After the value of $s$ has been updated, one resets the counter of $M$ back
to zero waiting for the next completed sweep to appear and repeats the process. 

Eventually, after several sweeps, once the gross features of $W$ have arisen,
$\delta W$ will get small since $s$ is being consistently lowered, and the value
of $M$ also should get smaller since fewer trajectories are needed for a completed
sweep to appear (that is the whole idea of this update). We consider the procedure
to be complete and $W(Q')$ to be ready for use in a Monte Carlo study when
$\delta W < 0.1$.  An animated GIF of how $W(Q')$ evolves in
this process is included in the Supplemental Material.

Our approach bears some similarities to the ``metadynamics'' approach
\cite{Barducci:2008zz,Laio:2015era,Sanfilippo:2017jyj}
which has also been considered for this problem.  One difference is in
the way the $W(\xi)$ function is found.  Some metadynamics
implementations vary $W(\xi)$ throughout the course of the evolution,
while others guess an initial value and keep it fixed.  We advocate a
hybrid approach where $W(\xi)$ is varied at first to optimize its
form, and then frozen to produce a truly detailed-balance respecting
evolution.  We also propose a specific, we believe quite efficient,
choice for the $W(\xi)$ function and its update.  The other difference is
that, in metadynamics, the $W(\xi)$ function is included as part of
a ``force'' term in an HMC evolution, whereas we implement it purely
through a Metropolis accept-reject step.  The force-term approach is more
efficient since HMC trajectories can be longer and because the
acceptance rate is higher.  But it requires evaluating the
field-derivative of $\xi$, which may not always be possible or
practical.  For instance, because we implement gradient flow through
stout smearing \cite{Morningstar:2003gk}, our $Q'$ is a differentiable
function of the link variables.  But because we use many stout
smearing steps, the differential expression is extremely unwieldy;
within a quenched simulation our Metropolis implementation is much
more efficient.  But because the fermionic force term is also very
expensive, the price may be worth paying in the unquenched case;
indeed, after our first draft of this paper appeared but before its final
publication, Bonati \textit{et al.} succeeded in applying such
a metadynamic method to the unquenched case \cite{Bonati:2018blm}.

Other approaches may also be available.  Recently Bonati \textit{et
al.} \cite{Bonati:2017woi} presented several 
algorithms to solve the problem of topological freezing
in the context of a simple quantum mechanical system which shares
basic similarities with the problem at hand. In particular, Sec.~IVF
contains very similar concepts as the ones used in this paper.
However it is not clear how the most effective algorithms they found
could be generalized to the topology problem in QCD.

\subsection{Parameters to Tune}
\label{subsec:totune}

The procedure described in the last section allows for the automated
determination of $W(Q')$, allowing for an efficient reweighting.  But
several parameters are still to be determined, and we found in
practice that a certain amount of hand tuning was needed to select
them.

First, there is the depth of gradient flow to use in establishing
$Q'$.  (In practice we actually used stout smearing
\cite{Morningstar:2003gk} with step-size 0.06 as our gradient flow
algorithm.  This would be totally inadequate if our goal were a
precision study of flowed operator expectation values, but here it is
only important to suppress UV fluctuations, so a more efficient if
less careful implementation of flow should be adequate.)
We found that $t'=0.24a^2$ is
insufficient to separate configurations of different topology, while
$t'=0.42a^2$ is enough; larger amounts of gradient flow start to destroy
the dislocations, which makes it more difficult to find the
configurations intermediate between topological sectors.  Optimally,
one should perform several beginning-to-end determinations of $\chi$,
each on the same lattice and temperature, but each time using different
$t'$ values, to do a systematic study of which choice leads to the
highest statistics for a fixed computational effort; but we have not
done this.

Second, there is the choice of the number and location of intervals.
We found 20 intervals to be adequate \textsl{except} that there were
two ``corners'' in the $W(Q')$ function where its slope rather
abruptly changes, see \figref{fig:Wfuncs}.  These appear to be the
points where the dominant type of configuration changes (regular
thermal configuration to dislocation, dislocation to full-sized
caloron), and the Monte Carlo simulation tends to get stuck at these
points.  We partly cured this by using more, narrower intervals at
these points, which handles finer structure in the reweighting
function and also leads to the algorithm spending more time near these
points.  So far we have done this by hand tuning, though presumably an
automated method of interval adjustment could be developed, based for
instance on the curvature of the determined $W[Q']$ function.

Next, there are the details of the parameter $s$ which controls how
fast we adjust the $W(Q')$ function.  We tried variations on the
procedure described above and found little change to the efficiency
with which a good $W(Q')$ is generated.  In any case, if high
statistics are desired, the Monte Carlo with fixed $W(Q')$ takes most
of the computational effort.

Next, there is the length of molecular-dynamics time used in the HMC
algorithm updates.  A larger HMC step leads to a larger change in the
configuration, which is good because it more efficiently explores the
phase space.  But it leads to larger changes in $Q'$ value and
therefore to a higher rejection rate.  So the HMC trajectory length
needs to be tuned to provide about 50\% acceptance rate in the
$\e^{\Delta W}$ acceptance step.  Again, 50\% is a rule of thumb; a
more careful analysis would compare the total achieved statistics at
fixed numerical effort as a function of HMC trajectory length.
To date we have not carried out such a study, and have instead used
the 50\% rule of thumb.  Our results in what follow used HMC
trajectory lengths of 0.2--0.25~$a$. We are well aware that such a
small trajectory length will result in big autocorrelation effects between
configurations. We have taken special care in providing a reliable error
estimate by making a careful error analysis based on binning and
jackknife \cite{Wolff:2003sm}.

Finally, there is the choice of the final observable used to determine
$\chi(T)$.  Every 100 HMC trajectories, we make a measurement of $Q$
which we use in our statistical analysis of the susceptibility.
We use $|Q|$ after some amount $t$ of gradient
flow and set its value to 1 if $|Q| \ge \Qthr$ and to 0 if
$|Q| < \Qthr$.  (Configurations with $|Q|>1$ are very rare
at the temperatures and for the volumes of interest, as we will
establish in the next section.) This leaves open the exact choice of
$t$, of $\Qthr$, and of gradient flow procedure
(Wilson versus Zeuthen \cite{Ramos:2015baa}).  All choices should lead
to the same continuum limit and it is not expensive to sample using
various choices and compare.  This is what we will do; any difference
between $\chi(T)$ values due to different threshold or flow depth will
indicate deficiencies in our lattice spacing, and must be seen to
vanish when we take the continuum limit.

\section{Results}
\label{sec: Results}

\begin{table*}
\centering
\caption{The lattices used in this work.  Those labeled with A
  correspond to $2.5~\Tc$ while the Bs are simulations at $4.1~\Tc$.}
\label{tab:ourlatts}
\begin{tabular*}{.85\textwidth}{l @{\extracolsep{\fill}} ccrrcc}
\hline\hline
Lat & $T/\Tc$ & $6/g^2_0$ &
$\beta/a$ & $L/a$ & \#Measurements & \#Complete sweeps
\\
\hline
$\text{A}_1$ & 2.5 & 6.507 \cite{Caselle:2018kap} & 6   & 16 & 61,759 & 591\\
$\text{A}_2$ & 2.5 & 6.722 \cite{Caselle:2018kap} & 8   & 16 & 96,068 & 263\\
$\text{A}_3$ & 2.5 & 6.903 \cite{Caselle:2018kap} & 10  & 24 & 66,840 & 195\\[.1cm]

$\text{B}_1$ & 4.1 & 6.883 \cite{Giusti:2010bb} & 6 & 16 & 70,699 & 313\\
$\text{B}_{2\mathrm a}$ & 4.1 & 7.135 \cite{Giusti:2010bb} & 8 & 8 & 50,992 & 94\\
$\text{B}_{2\mathrm b}$ & 4.1 & 7.135 \cite{Giusti:2010bb} & 8 & 12 & 50,390 & 82\\
$\text{B}_{2\mathrm c}$ & 4.1 & 7.135 \cite{Giusti:2010bb} & 8 & 16 & 52,900 & 145\\
$\text{B}_{2\mathrm d}$ & 4.1 & 7.135 \cite{Giusti:2010bb} & 8 & 24 & 74,900 & 168\\
$\text{B}_{2\mathrm e}$ & 4.1 & 7.135 \cite{Giusti:2010bb} & 8 & 32 & 72,800 & 151\\
$\text{B}_3$ & 4.1 & 7.325 \cite{Giusti:2010bb} & 10 & 24 & 82,663 & 104\\
\hline\hline
\end{tabular*}
\end{table*}

Our goal in this paper is to demonstrate our method and show that it
can obtain statistically powerful results at high temperatures in a
range of lattice spacings and volumes.  With this in mind, we study 10
different lattices, as listed in Tab.~\ref{tab:ourlatts}.  We use the
Wilson gauge action at two temperatures corresponding to $2.5~\Tc$
and $4.1~\Tc$; the values of $\beta$ are taken from Refs.~
\cite{Caselle:2018kap,Giusti:2010bb}.  At the higher
temperature we consider aspect ratios between 1:1 and 4:1 with
$N_\tau=8$ and at both temperatures we consider lattice spacings with
$N_\tau = 6,8,10$ with an aspect ratio of about 2.5:1.  This allows one
study of the volume scaling, and lets us take the continuum limit, but
it is not sufficient to consider both limits simultaneously.
All calculations were carried out over a six month period on one eight-core
desktop machine and one server node with two Xeon-Phi (KNL) CPUs.  By
modern standards this is an extremely modest computational budget.

\begin{figure}[t]
	\centering
	\includegraphics[width=1\linewidth]{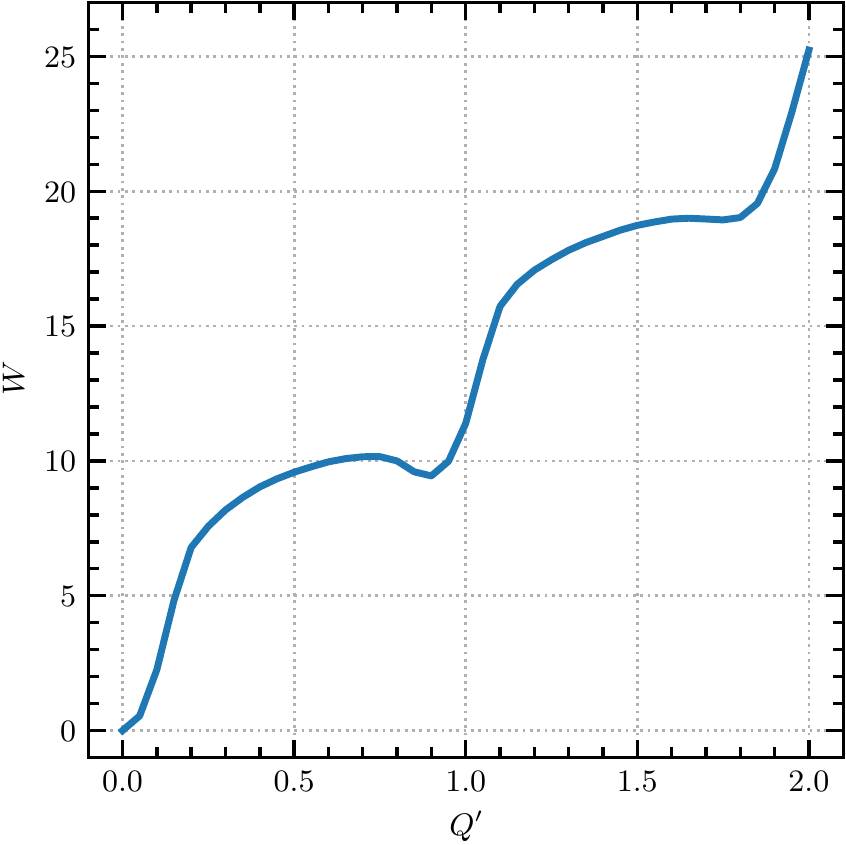}
	\caption{$W$ function of $6\times16^3$ lattice at $2.5~\Tc$
          (Lattice A$_1$).}
	\label{fig:WfuncQ2}
\end{figure}

The first question is:  is it sufficient so sample only $Q=0$ and
$|Q|=1$ sectors, or are larger values of $|Q|$ also important in
establishing the topological susceptibility?  To study this, first
look at \eqref{eq:exp_rew} and consider what happens when we use
$Q$ with a threshold as the observable:
\begin{align}
\begin{split}
  \label{doweneedQ=2}
  \langle Q^2 \rangle &= \frac{\sum_i^N Q_i^2 \e^{-W(Q'_i)}}{\sum_i^N
    \e^{-W(Q'_i)}}
  \\
  &\simeq \frac{\sum_{i:|Q|=1} \e^{-W(Q'_i)} + \sum_{i:|Q|=2} 4
    \e^{-W(Q'_i)} + \ldots}{\sum_{i:|Q|=0} \e^{-W(Q'_i)}},
    \end{split}
\end{align}
where in the numerator we only have to sum over $|Q|=1$ and higher
configurations since $Q=0$ does not contribute, while in the
denominator we only sum over $|Q|=0$ because they completely dominate
the ensemble.  Clearly we need both $Q=0$ and $|Q|=1$ configurations
to perform the calculation; but if our accuracy goal is $10\%$, then
we only need $|Q|=2$ and higher if the total probability to be in one
of these states is at least 2.5\% of that for $|Q|=1$ states.
Therefore we carried out the construction of $W(Q')$ in the domain
$0\le Q'\le2$, shown in \figref{fig:WfuncQ2}, for
the lower temperature we study and a $6\times 16^3$ lattice (Lattice
A$_1$).  We see immediately from the figure that
$Q' >1.5$ configurations require a reweighting of
$\e^{-18}$ to occur, while $Q'>0.75$ configurations occur
already with an $\e^{-9}$ reweighting.  For our $t'$ values the
$|Q|=1$ values all have $Q'>0.7$ and $|Q|=2$ values all
have $Q'>1.5$,
so this means that $|Q|=2$ configurations are suppressed relative to
$|Q|=1$ configurations by about $\e^{-9}$.  Therefore $|Q|=2$ plays a
tiny role in \eqref{doweneedQ=2} and can be safely ignored.
In a larger volume, an instanton gas estimate says that the $|Q|=1$
configurations should get more common with $V$ and the $|Q|=2$
configurations should get more common with $V^2$.  So $|Q|=2$ would
start to become relevant in a box with an aspect ratio of about 15.  Such
enormous lattices are not needed to study $\chi(T=2.5~\Tc)$,
and so we do not need to consider $|Q|\geq 2$.
This conclusion only strengthens for a larger $T$
where the susceptibility is still smaller.  Obviously, at some lower
temperature it will break down and we will need many topological
sectors; so every time we go to lower temperatures we must revisit
this issue.

\begin{figure}[t]
	\centering
	\includegraphics[width=1\linewidth]{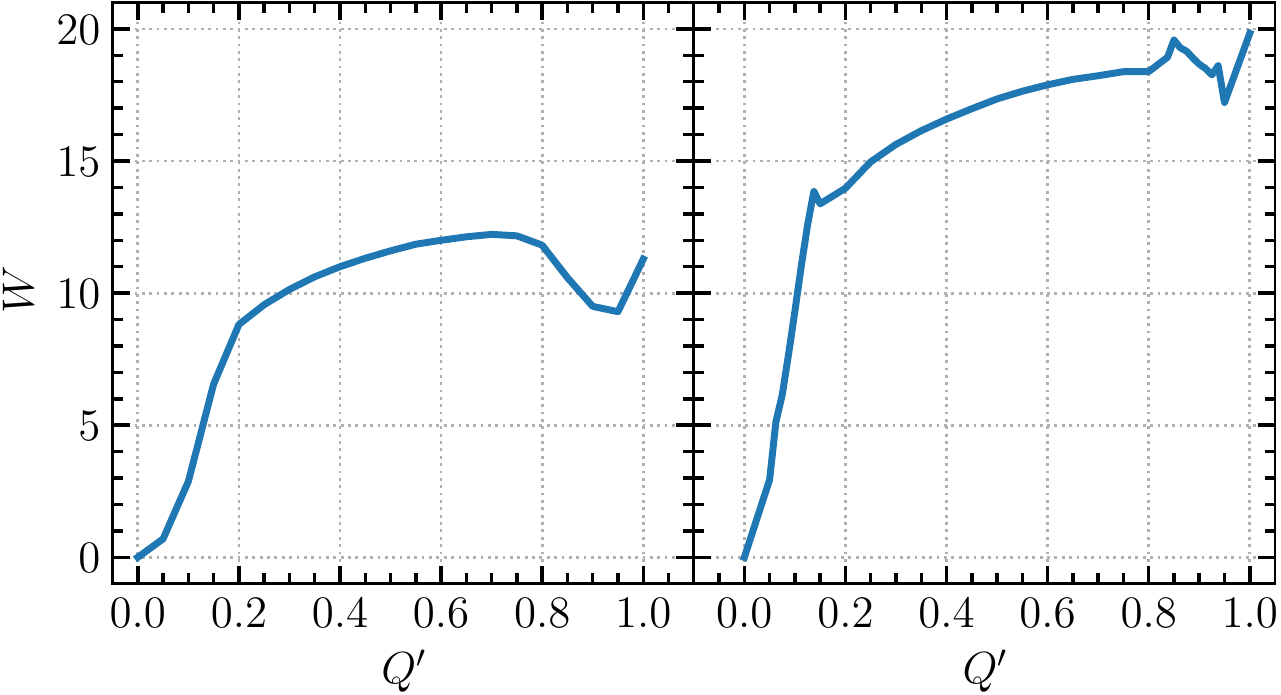}
	\caption{$W$ functions of $8\times16^3$ lattice. Left:
          $2.5~\Tc$ (Lattice A$_2$). Right: $4.1~\Tc$ (Lattice B$_{\mathrm{2c}}$).}
	\label{fig:Wfuncs}
\end{figure}

We proceed to compute the reweighting function $W(Q')$ using
$\Qmax'=1$.  Two examples are shown in \figref{fig:Wfuncs}.  In
each case there is a deep minimum at $Q'=0$ corresponding to ordinary
$Q=0$ configurations and a much shallower minimum near $Q'=1$,
corresponding to $\left|Q\right|=1$ configurations.  The broad plateau in between
can be understood as configurations containing a dislocation.  The
sharp features in the $4.1~\Tc$ plot are caused by our abruptly
adjusting the width of our intervals.  At finer lattices (larger
$N_\tau$) the $\left|Q\right|=1$ minimum becomes deeper (or more
accurately, the barrier gets higher), as the size of physical
calorons becomes more different from the lattice spacing.

\begin{table*}[htp]
\caption{This table shows $-\ln\left(\chi/\Tc^4\right)$ for all points plotted
in Figs.~\ref{fig:first_results}--\ref{fig:extrap2}.  Errors are statistical only.}
\label{tab:plotpoints}
\centering
\begin{tabular*}{\textwidth}{l @{\extracolsep{\fill}} cccccc}
\hline\hline
Flow: & Wilson & Wilson & Zeuthen & Zeuthen & Zeuthen & Zeuthen\\\hline
$t/a^2$: & $1.2$ & $2.4$ & $1.2$ & $2.4$ & $2.4$ & $2.4$\\
$\Qthr$: & 0.7 & 0.7 & 0.7 & 0.5 & 0.7 & 0.9\\[.1cm]
\hline
\hline
Lat. & $-\ln\left(\chi/\Tc^4\right)$ & $-\ln\left(\chi/\Tc^4\right)$ & $-\ln\left(\chi/\Tc^4\right)$ &
$-\ln\left(\chi/\Tc^4\right)$ & $-\ln\left(\chi/\Tc^4\right)$ & $-\ln\left(\chi/\Tc^4\right)$ \\[.1cm]
\hline
$\text{A}_1$ & 7.37(07) & 7.52(07) & 7.24(07) & 7.31(07) & 7.35(07) & 7.53(07)\\
$\text{A}_2$ & 7.79(10) & 7.85(10) & 7.74(10) & 7.76(10) & 7.78(10) & 7.85(10)\\
$\text{A}_3$ & 8.09(16) & 8.11(16) & 8.07(16) & 8.08(16) & 8.08(16) & 8.11(16)\\[.1cm]
$\text{B}_1$ & 10.21(07) & 10.41(07) & 10.04(07) & 10.14(07) & 10.19(07) & 10.43(07)\\
$\text{B}_{2\mathrm a}$ & 12.74(09) & 13.11(10) & 12.47(10) & 12.65(09) & 12.75(10) & 13.17(10)\\
$\text{B}_{2\mathrm b}$ & 11.90(10) & 12.08(11) & 11.74(10) & 11.84(10) & 11.89(11) & 12.10(11)\\
$\text{B}_{2\mathrm c}$ & 11.36(11) & 11.46(12) & 11.26(11) & 11.31(11) & 11.34(11) & 11.46(12)\\
$\text{B}_{2\mathrm d}$ & 11.10(13) & 11.18(13) & 11.03(13) & 11.07(13) & 11.09(13) & 11.18(13)\\
$\text{B}_{2\mathrm e}$ & 11.16(14) & 11.24(14) & 11.07(13) & 11.12(14) & 11.15(14) & 11.24(14)\\
$\text{B}_3$ & 11.76(17) & 11.80(17) & 11.72(17) & 11.74(17) & 11.76(17) & 11.80(17)\\[.1cm]
\hline\hline
\end{tabular*}
\end{table*}

\begin{figure}[htb]
    \centering
    \includegraphics[width=1\linewidth]{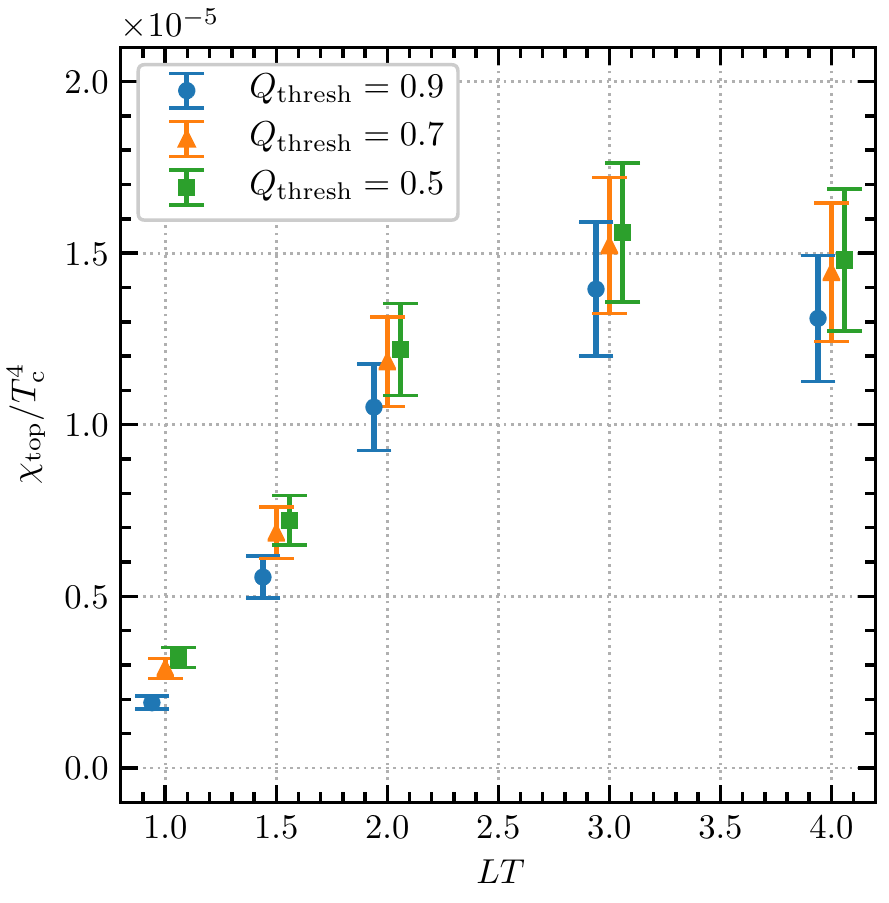}
    \caption{Finite volume dependence of the susceptibility at
      $N_\tau=8$ and $T = 4.1~\Tc$ (Lattices B$_{\mathrm{2a}}$ through
      B$_{\mathrm{2e}}$), using Zeuthen flow with $t=2.4 a^2$ and three
      different values of $\Qthr$ (points have been displaced for
      reasons of visibility).}
    \label{fig:first_results}
\end{figure}

With the $W(Q')$ reweighting functions in hand, we proceed
to evaluate the topological susceptibility via \eqref{doweneedQ=2}.  For completeness, we
present all of our results in Tab.~\ref{tab:plotpoints}.  The errors
in the table always represent our statistical uncertainty, for the
given lattice spacing, temperature, volume, and $Q$ definition.  Systematic
errors, particularly those associated with the continuum and
large-volume limits, must be determined by comparing results from
different lattices.
First, consider the large-volume limit, by analyzing $\chi$ as a function of
aspect ratio, shown in \figref{fig:first_results}.
The figure evaluates $Q$ after $t=2.4 a^2$ of improved (Zeuthen)
gradient flow, and considers three different values for the threshold
to distinguish between $|Q|=1$ and $Q=0$:  $\Qthr = 0.5$, 0.7, and 0.9.
The figure shows that, as expected, aspect ratios smaller than 2 are
badly discrepant; but the difference between an aspect ratio of 2 and
4 is of an order of tens of a percent, and is not statistically significant.
It appears that large-volume behavior sets in at a modest aspect ratio
between 2 and 3 (at this temperature).
Therefore in this exploratory study, we will only consider the
continuum limit for an aspect ratio of about 2.5.  The figure also
shows that although the value of $\Qthr$ introduces a systematic
effect (the lower the threshold, the higher the determined $\chi$ value),
this effect is statistically irrelevant already at $LT<1.5$ and
becomes even smaller for larger volumes (and finer lattices, see
Tab.~\ref{tab:plotpoints}); in what follows we will use $\Qthr = 0.7$.

\begin{figure*}[htp]
  \centering
   \includegraphics[width=1\linewidth]{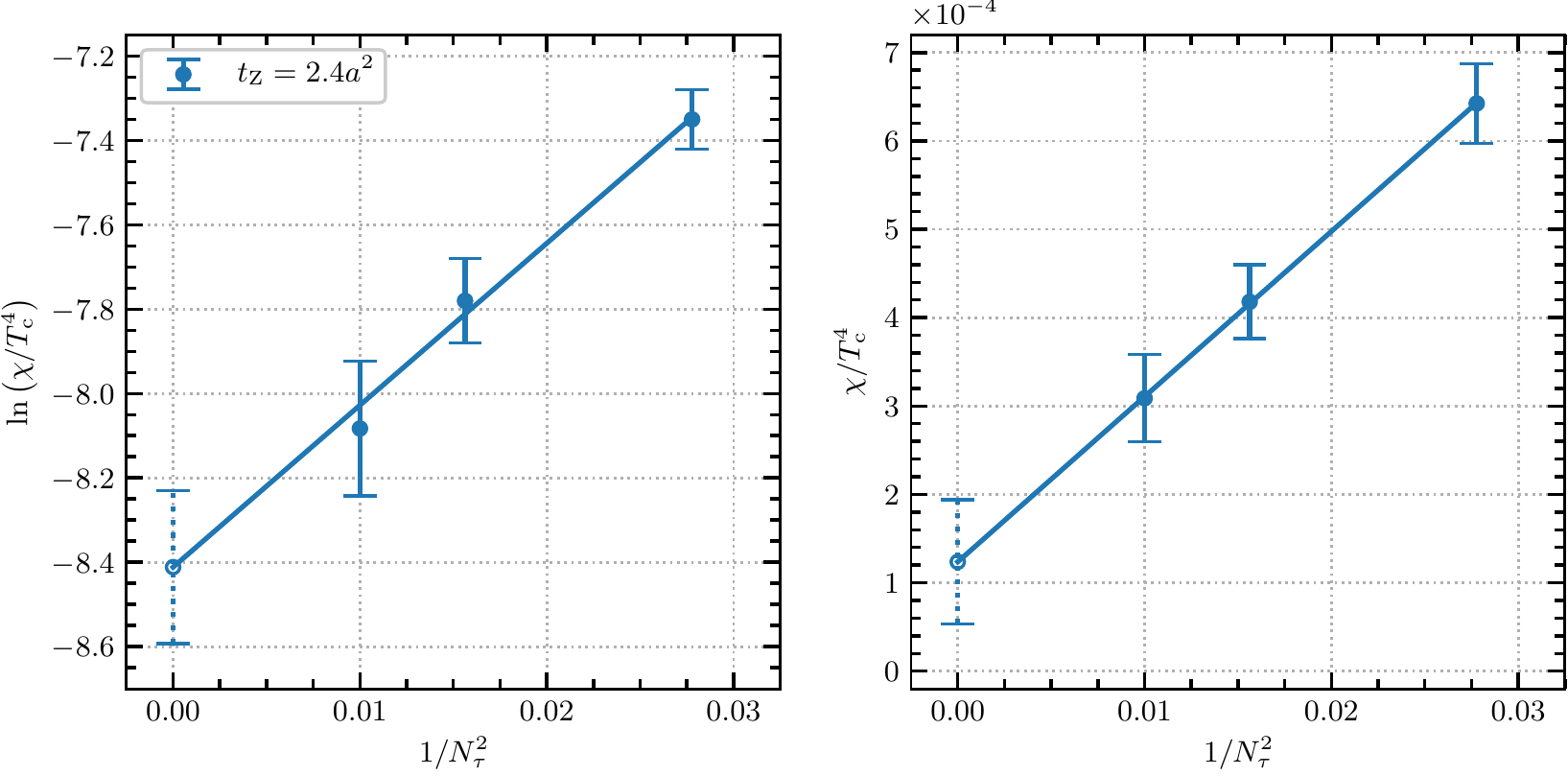}
  \caption{Continuum extrapolation at $T=2.5~\Tc$ based on lattices
    A$_1$, A$_2$, A$_3$, carried out in
    terms of $\chi(T)$ directly (right) and $\ln\left(\chi(T)\right)$
    (left) using Zeuthen flow with $t=2.4a^2$ and $\Qthr = 0.7$.}
  \label{fig:extrap1}
\end{figure*}

\begin{figure*}[htp]
  \centering
  \includegraphics[width=1\linewidth]{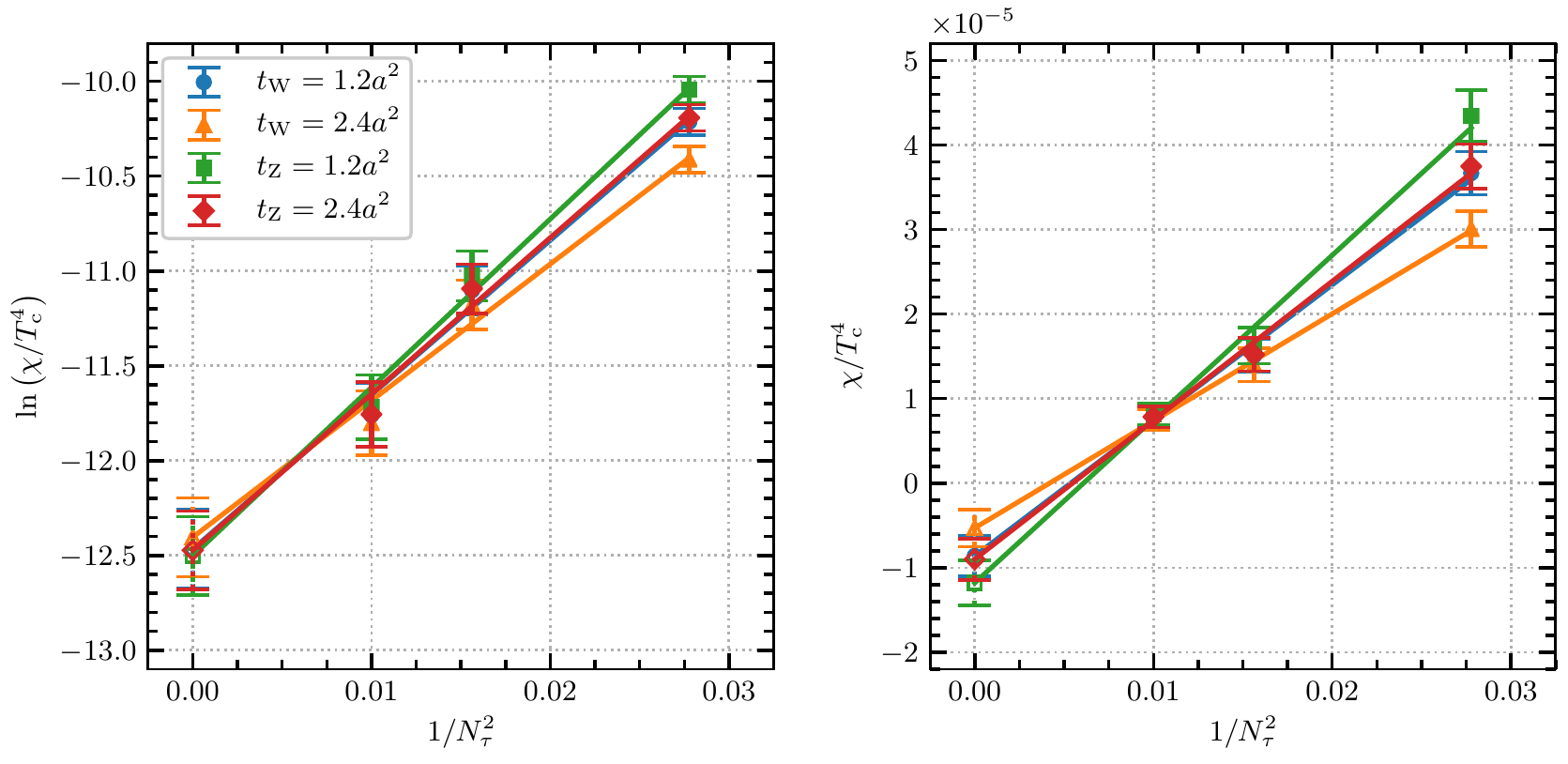}
  \caption{Same as \figref{fig:extrap1} except for $T=4.1~\Tc$, using
    lattices B$_1$, B$_{\mathrm{2d}}$, B$_3$.
  In addition we have shown separately the measured values for two
  amounts of flow $t=1.2 a^2$ and $t=2.4 a^2$ and for two flow
  actions (Wilson and Zeuthen). Note that the linear-extrapolated
  continuum limit is \textsl{negative} for all choices. }
  \label{fig:extrap2}
\end{figure*}

Finally we consider the continuum extrapolation, using three lattice
spacings with $N_\tau = 6,$ 8, and 10.  We show this for $T=2.5~\Tc$
in \figref{fig:extrap1} and $T=4.1~\Tc$ in
\figref{fig:extrap2} (note that the $N_\tau=8$ lattice has a slightly
different aspect ratio than the $N_\tau=6,10$ lattices; smaller for
$2.5~\Tc$ and larger for $4.1~\Tc$).  At the higher
temperature, we show results for
two flow depths ($t=1.2 a^2$ and $t=2.4 a^2$) and two choices of flow
action (Wilson and Zeuthen).  The different $Q$ definitions differ
significantly for $N_\tau=6$
(note that the determinations use the same Markov chain, so the errors
are highly correlated and the difference is statistically very
significant) but are nearly indistinguishable for $N_\tau=10$; so
issues of topology definition are seen to become small on fine
lattices.  The choice of topology definition is irrelevant in the
continuum limit \textsl{if} we extrapolate in terms of $\ln\left(\chi\right)$.

However, if we attempt to extrapolate $\chi\left(T,a^2\right)$ linearly against
$a^2$, we get very poor behavior. At $T=2.5~\Tc$ the two continuum
limits, based on extrapolating $\ln\left(\chi\right)$ and extrapolating $\chi$
directly, differ by more than
their error bars.  And at $T=4.1~\Tc$, the linear extrapolations of
$\chi(T)$ using different definitions of topology are  incompatible,
and each definition leads to a negative extrapolated value, which is
clearly unphysical.  On the other hand, if we perform a linear
extrapolation of $\ln\left(\chi\right)$ against $a^2$, the different definitions
of topology produce compatible results, which are finite and physical.
The reason that one should extrapolate in $\ln\left(\chi\right)$ and not in
$\chi$ directly, as we understand it
\cite{Jahn:2018jvx}, is that the topological susceptibility is
controlled by the exponential suppression of the caloron action
$\exp\left(-S_{\mathrm{caloron}}\right) = \exp\left(-8\pi^2 / g^2(\mu\sim T)\right)$.
This action receives multiplicative $\mathcal{O}\left(a^2\right)$ lattice
corrections:
$\chi \propto \exp(-S) \to \exp\left(-[1-\mathcal{O}(a^2 T^2)]\, S\right)$.
That is, the $a^2$ corrections are best viewed as a
shift in the caloron action and therefore in the logarithm of the
susceptibility.  Therefore an extrapolation of
$\ln\left(\chi(T)\right)$ in terms of $a^2$ is better justified, and
better behaved.  Indeed, \figref{fig:Wfuncs} shows that
$S_\mathrm{caloron}$ is about twice as large at $T=4.1~\Tc$ than at
$T=2.5~\Tc$; so the slope of the extrapolation should be twice as
large in the left panel of \figref{fig:extrap2} as in
\figref{fig:extrap1}, which it is.  Therefore this picture of the
nature of $a^2$ errors is consistent with our findings, and an
extrapolation of $\ln\left(\chi\right)$ against $a^2$ is the theoretically best
motivated way to extrapolate to the continuum.

\section{Discussion}
\label{sec: Discussion}

We have presented a methodology for applying reweighting
\cite{Berg:1991cf} to the measurement of topology in high temperature
pure-glue SU(3) QCD.  Our approach involves reweighting in terms of a
``poor man's'' topological measurement $Q'$ ($Q$ measured after a
small amount of flow $t'=0.42 a^2$ and using an $a^2$ improved
topological density operator).  There is then a two-stage simulation;
first we simulate while dynamically changing our reweight function, to
determine the form of the reweight function.  Then we fix the reweight
function and perform a Monte Carlo simulation to determine the
topological susceptibility.

The method is effective; with modest numerical
resources we are able to treat
$T=4.1~\Tc$ up to an aspect ratio of 4 and up to a lattice spacing
with $N_\tau = 10$, obtaining good statistics.  Making a full
continuum extrapolation but at a modest aspect ratio of 2.5
(extrapolating the $t=2.4a^2$, Zeuthen flow results), we find
\begin{align}
\begin{split}
  \frac{\chi(T=2.5~\Tc)}{\Tc^4} = 2.22 \times 10^{-4} \ \e^{\pm 0.18},
  \\
  \frac{\chi(T=4.1~\Tc)}{\Tc^4} = 3.83 \times 10^{-6} \ \e^{\pm 0.21}.
  \end{split}
\end{align}

Our results at individual $N_\tau$ values are consistent with previous
studies; our $\mathrm{A}_1$ lattice gives the same susceptibility as
found by Berkowitz \textsl{et al} \cite{Berkowitz:2015aua}, and our
results at $4.1\,\Tc$ and $N_\tau = 6,8$ (lattices $\mathrm{B}_{1}$
and $\mathrm{B}_{2\mathrm{e}}$) appear compatible with those at
$4.0\,\Tc$ from Borsanyi \textsl{et al} \cite{Borsanyi:2015cka},
who have significantly larger statistical errors despite applying much
more numerical effort.  Those authors also provide a
continuum-extrapolated functional fit for $\chi(T)$, which is in
reasonable agreement with our results; applying their fit form to the
temperatures we studied, we obtain
$\chi(2.5~\Tc)=1.9 \times 10^{-4}~\Tc^4$ and
$\chi(4.1~\Tc)=5.6 \times 10^{-6}~\Tc^4$.

Our results teach a few other lessons.  On a lattice with $N_\tau=6$,
$\chi$ is sensitive to the exact definition of topology (depth of
flow, flow action, threshold).  This dependence is nearly gone by
$N_\tau=10$ and seems not to affect the continuum extrapolation.  The
continuum extrapolation should be performed in terms of $\ln\left(\chi\right)$,
not in terms of $\chi$ itself.  The continuum extrapolation
corrections to $\ln\left(\chi\right)$ can be large and are
larger at higher temperature.  None of these lessons should be
surprising \cite{Jahn:2018jvx}.

\begin{figure}
  \centering
  \includegraphics[width=1\linewidth]{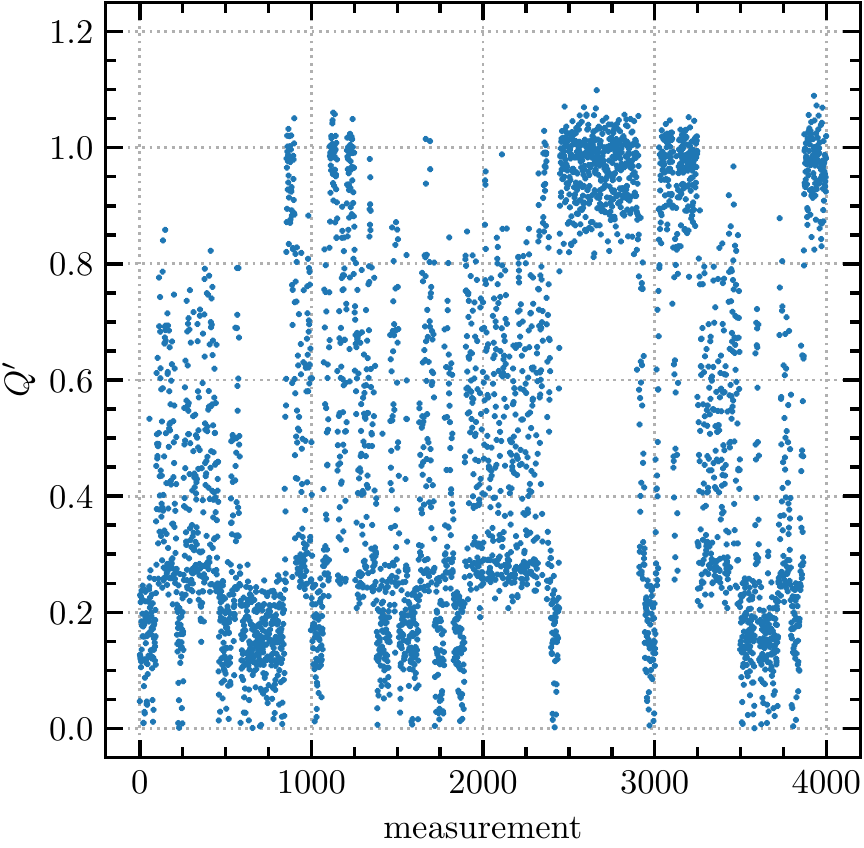}
  \caption{Piece of a Markov-chain history of $Q'$
    against measurement number, for lattice B$_{\mathrm{2d}}$.
    The reweighting allows efficient sampling
    in three regions, $0<Q'<0.27$, $0.20<Q'<0.85$,
    and $0.8<Q'<1.05$, but has difficulty moving between
    these regions.}
    \label{fig:chainbit}
\end{figure}

We should not claim that our technique solves all problems, however.
Looking at the $Q'$ value as a function of measurement number for a
short portion of a Markov chain evolution, shown in
\figref{fig:chainbit}, we see that despite our reweighting function, there
are a few points where the simulation gets ``stuck."%
\footnote{%
  Of course, without reweighting, not a single point in the plot would
  get above $Q'=0.15$.}
It moves easily in the range $0 <Q' < 0.27$
and similarly moves easily across $0.2 <Q' < 0.85$; but it has
difficulty moving from one of these ranges to the other.  There is a
similar ``barrier'' around $Q'= 0.8$.  These problems become more
severe as we move to larger $N_\tau$.  We
believe that this occurs because $Q'$ is an incomplete descriptor
which is missing some other information which distinguishes between
these regions.  We partly overcame this problem by making more,
narrower reweighting bins in these overlap regions; our reweighting
procedure causes the Markov chain to spend approximately equal time in
each bin, so narrower bins cause more time to be spent in these
regions, which helps the Markov chain to find the way between the
different regions.  (This is the reason for the cuspy discontinuities
in \figref{fig:Wfuncs}.)  However, while this helps, it hardly solves
the problem, as \figref{fig:chainbit} attests.
We are searching for one or more additional observables
to serve as further reweighting variables in the hopes of improving
this sampling.  Another inefficiency is that the number of updates
needed to move between topological sectors does not improve as we
increase the volume.  Therefore, to achieve a given level of
statistics, the numerical effort must grow linearly with the volume.
We do not foresee any solution to this problem.

Conceptually there are no obstacles to applying our technique to the
unquenched case (at high temperatures where only one or a few
topological sectors are relevant).  However, we expect doing so to be
numerically more
difficult.  First, the HMC algorithm requires far more computer power
with fermions.  Second, the $Q=1$ sector has near-zero eigenvalues, while
the $Q=0$ sector should have the smallest eigenvalue close to
$\pi T$.  The chiral limit should be severe.  Third, the
characteristic size of a caloron is smaller with light quarks than
without \cite{Gross:1980br}, and therefore the lattice spacing
should need to be smaller ($N_\tau$ values larger) than what we need
in pure glue.  But we view these added numerical challenges as reasons
that such studies should use reweighting.  The temperature range
where topology is relevant for axions is $3\,\Tc$ to $7\,\Tc$
\cite{Moore:2017ond}, where topology is quite suppressed and only the
$|Q|=1$ sector should contribute.  To overcome the challenges just
mentioned in this temperature range, we absolutely need the
improvement in statistical sampling of $|Q|=1$ from reweighting
if any statistical power is to be achieved.

It is less clear that our approach has applications at lower
temperatures where multiple topological sectors are relevant.
We might hope that a similar reweighting method might help with the
topological-sector sampling problem, which afflicts fine
lattices.  However, it is not clear to us that $Q'$ will be an
effective reweighting variable in this case.  We leave study of this
problem for future work.

\begin{acknowledgments}
We thank Kari Rummukainen, Edwin Laermann, Christof
Gattringer, Philippe de Forcrand, and Olaf Kaczmarek for valuable discussions.
The authors acknowledge support by the Deutsche Forschungsgemeinschaft
(DFG) through Grant No. CRC-TR 211,  ``Strong-interaction matter under
extreme conditions." We also thank the GSI Helmholtzzentrum and the TU
Darmstadt and its Institut f\"ur Kernphysik for supporting this
research. This work was performed using the framework of the publicly
available openQCD-1.6 package \cite{openQCD}.
\end{acknowledgments}

\bibliographystyle{apsrev4-1}
\bibliography{refs}

\end{document}